\documentclass[12pt]{iopart}

\usepackage{latexsym,amssymb, amsthm, epsfig}
\usepackage{graphicx} 
\usepackage{psfrag}  
\usepackage{graphics}

\newcommand{\Z}{{{\mathbb{Z}}}}
\newcommand{\R}{{{\mathbb{R}}}}
\newcommand{\N}{{{\mathbb{N}}}}
\newcommand{\E}{{{\mathbb{E}}}}

\def\P{{\mathbb P}}
\def\tsigma{{\tilde\sigma}}
\newcommand{\bsigma}{\mbox{\boldmath$\sigma$}}
\newcommand{\btsigma}{\mbox{\boldmath$\tilde\sigma$}}
\newcommand{\bxi}{\mbox{\boldmath$\xi$}}

\def\tX{{\tilde X}}

\def\ty{{\tilde y}}
\def\tC{{\tilde C}}
\def\tN{{\tilde N}}
\def\ts{{\tilde s}}

\def\by{\mathbf{y}}
\def\bs{\mathbf{s}}
\def\bts{\mathbf{\tilde s}}

\def\bty{\mathbf{\tilde y}}
\def\eqref#1{(\ref{#1})}
\def\one{{\mathbf 1}}  
\def\square{\ifmmode\sqr\else{$\sqr$}\fi}
\def\sqr{\vcenter{
         \hrule height.1mm
         \hbox{\vrule width.1mm height2.2mm\kern2.18mm\vrule width.1mm}
         \hrule height.1mm}}                  

\theoremstyle{plain}   
\newtheorem{teo}{Theorem}
 
\newtheorem{propo}[equation]{Proposition}

\newtheorem{vthm}[equation]{Lemma}

\begin{document}

\title[A slow-to-start traffic model related to a M/M/1 queue] {A slow-to-start
  traffic model\\ related to a M/M/1 queue}

\author{Fredy Castellares C\'aceres$^1$, Pablo A.  Ferrari$^2$, \\
Eugene Pechersky$^3$}

\address{$^1$ Departamento de Estat\'istica, Universidade Federal de Minas
  Gerais, 31270-901 Belo Horizonte-MG. Brasil. }

\address{$^2$ Instituto de Matem\'atica e Estat\'{\i}stica, Universidade de
  S\~ao Paulo, Rua do Mat\~ao 1010, 05508-090 S\~ao Paulo, Brasil.}

\address{$^3$ IITP, 19, Bolshoj Karetny per., GSP-4 Moscow 127994, Russia. }

\ead{fredy@est.ufmg.br, pablo@ime.usp.br, pech@iitp.ru}

\begin{abstract}
  We consider a system of ordered cars moving in $\mathbb R$ from right to left.
  Each car is represented by a point in $\R$; two or more cars can occupy the
  same point but cannot overpass. Cars have two possible velocities: either $0$
  or $1$. An unblocked car needs an exponential random time of mean 1 to pass
  from speed 0 to speed $1$ (\emph{slow-to-start}). Car $i$, say, travels at
  speed $1$ until it (possibly) hits the stopped car $i-1$ to its left. After
  the departure of car $i-1$, car $i$ waits an exponential random time to change
  its speed to $1$, travels at this speed until it hits again stopped car $i-1$
  and so on. Initially cars are distributed in $\R$ according to a Poisson
  process of parameter $\lambda<1$. We show that every car will be stopped only
  a finite number of times and that the final relative car positions is again a
  Poisson process with parameter $\lambda$. To do that, we relate the
  trajectories of the cars to a $M/M/1$ stationary queue as follows.  Space in
  the traffic model is time for the queue. The initial positions of the cars
  coincide with the arrival process of the queue and the final relative car
  positions match the departure process of the queue.
\end{abstract}
\pacs{1315, 9440T}
\maketitle

\section{Introduction}

Interest in traffic models is old. In 1935 Greenshield (see Helbing
\cite{Helbing} and Chowdhury \emph{et al} \cite{CSS} for a review on traffic
models) introduced the first traffic study. In 1959 Greenberg \cite{Greenberg}
called the attention to the importance of the area.  In 1992 Nagel-Schreckenberg
\cite{Nagel} study traffic using probabilistic cellular automata, computer
simulations and mean field models. \emph{Slow-to-start} models intend to capture
the behavior of cars that come out of a traffic jam: a driver needs a moment to
speed the car. Nagel and Schreckenberg \cite{Nagel}, Schadschneider and
Schreckenberg \cite{SS}, Gray and Griffeath \cite{GG}, Chowdhury, Santen and
Schadschneider \cite{CSS} and Yaguchi \cite{Y} studied these type of
models. Other studies of traffic have been done by Krug and Spohn \cite{KS},
Barlovic, Santen, Schadschneider and Schreckenberg \cite{Barlovic1}, Fuks
\cite{Fuks1}, Boccara and Fuks \cite{Boccara}, Belitsky and Ferrari\cite{BF2},
Belitsky, Krug, Neves and Sch\"utz \cite{Bel}, Blank \cite{Blank1,Blank2}, Helbing
\cite{Helbing}, Wolfram \cite{W} among many others.

We introduce a slow-to-start traffic model which is continuous in space and
time. Initially cars are distributed in the straight line $\mathbb{R}$ following
a Poisson process of parameter $\lambda$; cars may have speed $1$ or 0. All cars
start at zero speed and each car waits a random (delay) time exponentially
distributed with mean $1$ to to change its speed from 0 to 1.  Delay times of
different cars are independent. After the delay time, each car moves at speed
$1$ until it collides with a stopped car to its left or forever if it does not
collide. When car $i$ collides with car $i-1$, its speed drops immediately to
zero and remains blocked until car $i-1$ leaves.  Then car $i$ waits a further
random time with exponential distribution before moving. And so on.  At each
time there are cars at speed $1$ and cars at speed $0$. We prove that if
$\lambda<1$ every car will eventually be unblocked forever and the final
relative positions of cars are distributed as a Poisson process of rate
$\lambda$.

The main tool is a relation between the traffic model and a $M/M/1$ stationary
queuing system. Customers arrive at rate $\lambda$ according to a Poisson
process to a single server whose service times are independent and exponentially
distributed with mean $1$. There exists a stationary version for the queue when
$\lambda<1$. The stationary process is reversible so that its distribution is
invariant by reversing the arrow of time. As a consequence the customer
departure process is a Poisson process as the arrival process. This is the
famous Burke's theorem.

In Section \ref{S2} we define the process and give the main results. In Section
\ref{S3} we construct a semi-infinite traffic model and define traffic
cycles. In Section \ref{S4} we define the queue process and construct workload
cycles. In Section \ref{S5} we relate traffic final relative car-positions and
customer exit times in the queue. In Section \ref{S6} we use an approach of
Thorisson to construct the stationary traffic process using cycles and in
Section \ref{S7} we do it for the queue and conclude the proofs of the results.

\section{Definition of model and main results}
\label{S2}

We consider a system of cars moving in $\R$ from right to left. Cars have two
possible velocities: either $0$ or $1$. Cars are represented by points and
cannot overpass. A car needs an exponential random time of mean 1 to pass from
speed 0 to speed $1$. A typical car starts at speed 0, waits an exponentially
distributed random time to change its speed to $1$, travels at this velocity
until it is blocked by another stopped car, waits the stopped car to leave,
waits another exponential time to get again velocity $1$ and so on.

We construct a sequence $(\Pi,V)= ((\Pi(t),V(t));\,t\ge 0)$ of car trajectories,
$\Pi(t)=(\pi_i(t), i\in \Z)$ and $V(t)=(v_i(t), i\in \Z)$. For each $i$,
$\pi_i:\R_+ \to \R$ is a piecewise linear function almost everywhere
differentiable; $\pi_i(t)$ represents the position of car $i$ at time $t$ and
$v_i(t)\in\{0,1\}$ its velocity. The initial car positions are given by
$(y_i,\,i\in\Z)$, with $y_i\in\R$ and $y_i<y_{i+1}$ for all $i$: set
$\pi_i(0)=y_i$. The initial velocities are all null: $v_i\equiv 0$.  The
trajectories must satisfy the following properties:
\begin{enumerate}
\item $\pi_i(0)=y_i$
\item $\dot\pi_i(t) = -v_i(t)$ if $v_i$ is continuous at $t$
\item $v_i(t)$ jumps from $0$ to $1$ at rate 1 if $\pi_i(t)>\pi_{i-1}(t)$.
\item $v_i(t) =0$ if  $\pi_i(t)= \pi_{i-1}(t)$
\end{enumerate}

If instead of taking $i$ in $\Z$ we consider a semi-infinite configuration of
cars with initial positions $y_0\le y_1\le\dots$, the trajectories can be
constructed inductively.  Car 0 waits an exponential time of rate 1 and then
assumes velocity $1$; since it will not be blocked by any other car, it will
continue at speed $1$ for ever. Car 1 does the same, but if it collides with car
0 (at $y_0$), then it stops and after car 0 leaves, it waits an extra
exponential time and then goes, and so on. Let $\pi_i := (\pi_i(t), \,t\ge
0)$. If we think $i$ as discrete time, the process of trajectories is Markovian
in the sense that the law of the trajectory $\pi_i$ given
$\pi_{i-1},\dots,\pi_0$ depends only on $\pi_{i-1}$. Our first result says that
it is possible to construct a spatially translation invariant traffic process if
the initial positions of the cars are given by a Poisson process of rate
$\lambda$.

\begin{propo}
\label{p33}
If $(y_i;\,i\in\Z)$ is a stationary Poisson process in $\R$ with rate $\lambda$
and $\lambda<1$ then there exists a spatially stationary version of the process
$\Pi=((\pi_i(t);\,t\ge 0),\,i\in\Z)$ with initial positions $\pi_i(0)=y_i$ for
all $i\in\Z$.
\end{propo}

In the stationary Poisson process by convention $y_0<0<y_1$, so that $-y_0$,
$y_1$ and $y_{i+1}-y_i$ for $i\neq 0$ are iid random variables exponentially
distributed.  Let \[t_i := \sup\{s:\, v_i(s)=0\}\] (possibly equal to infinity)
be the last time car $i$ has velocity 0. From $t_i$ on, car $i$ goes freely at
speed $1$.  We say that car $i$ is \emph{free} at times $t>t_i$.

\begin{propo}
\label{t6}
Assume $\lambda<1$ and consider the traffic process $\Pi$ of Proposition \ref{p33}. For
each $i\in\Z$, $t_i$ is finite almost surely.
\end{propo}

Let $D_i$ be the \emph{total delay} of car $i$ defined by 
\begin{equation}
  \label{63}
  D_i := |-t_i-(\pi_i(t_i)-y_i)|
\end{equation}
Here $-t_i$ is the displacement car $i$ would have at time $t_i$ if it started
at speed $1$ and had never been blocked and $\pi_i(t_i)-y_i$ is the actual
displacement by that time. The absolute value of the difference is the delay of
car $i$. Call
\begin{equation}
  \label{s_i}
  s_i := y_i+D_i
\end{equation}
To understand the meaning of $s_i$, observe that if a car starts at position
$s_i$ at speed $1$ and it is never blocked, then its position at time $t$
coincides with $\pi_i(t)$ for all times $t\ge t_i$.

\begin{propo}
\label{t6a}
  If $\lambda<1$ then $\{s_i;\,i\in\Z\}$ is a Poisson process of parameter
  $\lambda$. Furthermore, $\pi_i(t) -\pi_{i-1}(t)= s_i-s_{i-1}$ for $t > \max\{t_i, t_{i-1}\}$.
\end{propo}
The statement is that $\{s_i;\,i\in\Z\}$ \emph{as a subset of $\R$} is a Poisson
process; the specification of the indexes may corrupt the Poisson property.
These results say that if the initial position of cars are distributed according
to a Poisson process and all with zero speed, then every car will eventually
have velocity $1$ and the relative car positions will be distributed according to
a Poisson process. For this reason we call $s_i$ the \emph{final relative
  position} of car $i$.

Let $y\in\R$ and for initial position $y_i>y$ let $r_i$ be the random time defined by
\[
r_i(y) := \sup ((\pi_i)^{-1}(y))
\]
this is last time the time car $i$ is at position $y$.
Let $T(y)$ be the first time all cars crossing $y$ after $T(y)$ are free:
\[
T(y):= \inf \{r_i(y);\, t_i\le r_m(y) \hbox{ for all } m\ge i\}
\]

\begin{propo}
\label{p111}
  If $\lambda<1$ then $T(y)<\infty$ almost surely for all $y\in\R$.
\end{propo}

\section{Construction of trajectories for a semi-infinite initial car
  configuration}
\label{S3}
We construct car trajectories for initial car positions $0=y_0 < y_1<\dots$ and
relate them to a $M/M/1$ queue starting with an empty system at arrival of a
customer.

The trajectories will be defined as a function of a marked Poisson process
\begin{equation}
  \label{p34}
  (\by,\bxi)=((y_i,\xi_{i,m});\ i\ge 0,$ $ 0\le m\le i)\,
\end{equation}
where $\by=(y_i, i\ge 0 )$ is a Poisson process on $[0,\infty)$ with rate
$\lambda$ with a car added at the origin (that is, $y_0=0$ and $(y_{i+1}-y_i ;
i\ge 0)$ are \emph{iid} exponential random variables with mean $1/\lambda$) and
the \emph{delay times} $\bxi=((\xi_{i,m} ; \ 0\le m\leq i), i\ge 0)$ are
\emph{iid} exponential random variables with mean~$1$.  The sequences $\bxi$ and
$\by$ are independent. These times are used as follows. Car $i$ may collide with
car $i-1$ at sites $y_0,\dots,y_{i-1}$; if the collision occurs at $y_m$, then
after car $i-1$ leaves $y_m$, car $i$ waits $\xi_{i,m}$ units of time before
taking again speed $1$.  More rigorously, car 0 starts at the origin and waits
$\xi_{0,0}$ units of time to start moving at speed $1$. Since there are no cars
to the left of car 0, it will never be blocked and we define
\begin{equation}
  \pi_0(t):=
  \left\{
  \begin{array}{ll}
    0& \mbox{if}\,\ 0<t\leq\xi_{0,0}\\
    -t+\xi_{0,0} & \mbox{if} \,\ t>\xi_{0,0}
  \end{array}
\right.
\end{equation}
The trajectory of car $i$ is then defined as a function of the trajectory of car
$i-1$ and the waiting times $(\xi_{i,m};\, 0\le m\le i)$ as follows. We shall
define $A_{i,m}$ as the time car $i$ arrives to $y_m$ and $B_{i,m}$ as the time
car $i$ departs from $y_m$.  Clearly $A_{i,m}\le B_{i,m}$ and if car $i$ is not
blocked at $y_m$, then $A_{i,m}=B_{i,m}$.  Let
\[
A_{i,i}:=0, \quad B_{i,i} := \xi_{i,i}
\]
and then inductively assume $A_{i-1,m}, B_{i-1,m}$ are defined for all $0\le
m\le i-1$, as well as $A_{i,m}$ and $B_{i,m}$. Then set
\begin{eqnarray}
  \label{t33}
  A_{i,m-1}:=B_{i,m} + y_m-y_{m-1}\\
  B_{i,m-1}:=\left\{
    \begin{array}{ll} A_{i,m-1}& \mbox{if}\,\ A_{i,m-1}> B_{i-1,m-1}\\
      B_{i-1,m-1}+\xi_{i,m-1} & \mbox{if} \,\ A_{i,m-1}< B_{i-1,m-1}
    \end{array}\right.
\end{eqnarray}
In words: the arrival of car $i$ to site $m-1$ occurs $(y_m-y_{m-1})$ time units
after its departure from site $m$. The departure of car $i$ from site $y_{m-1}$
occurs immediately (at arrival time) if car $i-1$ has already left or,
$\xi_{i,m-1}$ units of time after $B_{i-1,m-1}$, the departure of car $i-1$ from
site $y_{m-1}$.

The vector $((A_{i,m};\,B_{i,m}),\, 0\le m\le i)$ determines the trajectory
$(\pi_i(t), t\ge 0)$:
\begin{equation}
  \label{t33a}
  \pi_i(t) := \sum_{k=0}^i y_k \one\{t\in (A_{i,k}, B_{i,k})\} +(y_k+B_{i,k} -t) \one\{t\in
  (B_{i,k}, A_{i,k-1}) \}
\end{equation}
where $\one \{\cdot\}$ is the indicator function of the set $\{\cdot\}$.  The
total delay of car $i$ defined in \eqref{63} satisfies
\begin{eqnarray}
  D_i = \sum_{k=0}^i (B_{i,k}-A_{i,k})
\end{eqnarray}
To stress the dependence on $(\by,\bxi)$ we write $A_{i,k}(\by,\bxi)$,
$B_{i,k}(\by,\bxi)$, etc.
Let $\bs(\by,\bxi)=(s_0,s_1,\dots)$, where the final relative position $s_i$ is
defined as a function of $y_i$ and $D_i$ as in~\eqref{s_i}.
Let $\bsigma(\by,\bxi)=(\sigma_0,\sigma_1,\dots)$ be the sequence defined by
$\sigma_0:=\xi_{0,0}$ and for $i\ge 1$
\begin{equation}
  \label{t2}
  \sigma_i := \left\{\begin{array}{ll} 
      B_{i,0}-B_{i-1,0}& \mbox{if}\,\ s_{i-1}>y_{i}, \\
      \xi_{i,i} & \mbox{otherwise} 
    \end{array}\right.
\end{equation}
$\sigma_i$ is called the \emph{final delay} of car $i$.

\begin{vthm}
\label{t66}
$\bsigma=(\sigma_0,\sigma_1,\dots)$ is a sequence of iid random variables with
exponential law of mean~1. Furthermore $\bsigma$ is independent of $(y_i;\, i\ge
0)$.
\end{vthm}

\paragraph{Proof}
Fix the trajectory of car $i-1$. There are two cases: either car $i$ is blocked
at 0 by car $i-1$ or not. In the first case $B_{i,0}=B_{i-1,0}+\xi_{i,0}$, so
that $\sigma_i=\xi_{i,0}$ which is independent of the trajectories $(\pi_{m};\,
0\le m\le i-1)$ and in particular of $(\sigma_m;\,0\le m\le i-1)$. In the second
case the label of the leftmost blocking position of car $i$ is given by
\[
K := \min\{k\le i\,:\, B_{i-1,k}+\xi_{i,k} > B_{i-1,0}-(y_k-y_0) \}
\]  
here by convention $B_{i-1,i}=0$.  $K$ is a stopping time for $(\xi_{i,i-m};\,
m\ge 0)$; that is, the event $\{K=k\}$ is a function of $(\xi_{i,i-m};\, 0\le
m\le k)$. But the dependence on $\xi_{i,k}$ is only on the event $\{\xi_{i,k} >
B_{i-1,0}-(y_k-y_0)-B_{i-1,k}\}$.  Then, given this event,
\[
\sigma_i = \xi_{i,K} - ( B_{i-1,0}-(y_K-y_0)-B_{i-1,K})
\]
is exponentially distributed with mean one and independent of $(\pi_{m};\, 0\le
m\le i-1)$.

\paragraph{\bf Traffic Cycle}
Call
\begin{eqnarray}
  \label{p18}
  X:= \min\{y_i>0\,:\, y_i>s_{i-1}\}\\
  N := \min\{i>0\,:\,  y_i>s_{i-1}\}\\
  C := \big(((\pi_i(t);\,t\ge 0),\, i\in\{0,\dots,N-1\}),N,X\big)
\end{eqnarray}
We say that $C$ is a \emph{cycle} with \emph{length} $X$ and $N$ cars involved.
The cycle $C$ consists of a space interval $X$ and $N$ car trajectories in the
time interval $[0,\infty)$ with starting positions in $[0,X)$; however, since
car $i$ is free after time $B_{i,0}$, the trajectories are determined by the set
$((\pi_i(t);\,t\in[0,B_{i,0}));\, i\in\{0,\dots,N-1\})$ or, alternatively by the
arrival/departure times of the $N$ cars to/from sites $y_0,\dots,y_{N-1}$ given
by $((A_{i,m},B_{i,m});\, 0\le i\le N-1, 0\le m \le i)$.

The cycle $C$ induces a stochastic process $Z(C)=(Z_y(C), \, y\in[0,X))$ given
by the interval-valued vector (with dimension depending on $y$)
\begin{equation}
  \label{p24}
  Z_y(C) := (\pi_i^{-1}(y);\,y\le y_i<X)
\end{equation}
The $k$th coordinate of $Z_y(C)$ contains the time interval spent at $y$ by the
$k$th car to the right of $y$. Assuming this car has label $i$ there are two
cases: (a) if $y=y_m$ for some $m$, the interval is $\pi_i^{-1}(y)=
[A_{i,m},B_{i,m}]$ and (b) if $y$ is not an initial car position the interval is
a point, because car $k$ will not be delayed at $y$. $Z_y(C)$ is a vector of
zero length if $y\in[y_{N-1},X)$.

The cycle $C$, its length $X$ and its number o cars $N$ are functions of $\by =
(y_0,y_1,\dots)$ and $\bxi = (\xi_{i,m};\, i\ge 0,\, 0\le m \le i)$:
\[
C = C(\by,\bxi)\,;\quad X = X(\by,\bxi)\,;\quad N = N(\by,\bxi)
\]
Given a cycle $C$ we can recover the length of the cycle, the number of cars
involved, the initial car positions, the final delay
times and the final relative car positions which are denoted
\begin{eqnarray}
  \label{p38}
 N(C)\,,\; X(C)\,;\quad
 y_i(C),\; \sigma_i(C),\; s_i(C),\quad i=0,\dots,N(C)-1\,.
\end{eqnarray} 
The final relative car positions in the cycle coincide with the last passage
through the origin:
\begin{equation}
  \label{p51}
  s_i = B_{i,0},\quad \hbox{for }i=0,\dots,N-1
\end{equation}

\begin{figure}[!htp]

\psfrag{k1}{$t$}
\psfrag{k2}{$y$}
\psfrag{k3}{$y_{0}$}
\psfrag{k4}{$y_{1}$}
\psfrag{k5}{$y_{2}$}
\psfrag{k6}{$s_{0}$}
\psfrag{k7}{$s_{1}$}
\psfrag{k8}{$s_{2}$}
\psfrag{k9}{$y_{4}$}
\psfrag{k10}{$s_{4}$}
\psfrag{k20}{$s_{3}$}
\psfrag{k21}{$y_{3}$}

\psfrag{k11}{$\pi_{0}$}
\psfrag{k12}{$\pi_{1}$}
\psfrag{k13}{$\pi_{2}$}
\psfrag{k14}{$\pi_{4}$}
\psfrag{k19}{$\pi_{3}$}

\psfrag{k15}{$\sigma_{0}$}
\psfrag{k16}{$\sigma_{1}$}
\psfrag{k17}{$\sigma_{2}$}
\psfrag{k18}{$\sigma_{4}$}
\psfrag{k22}{$\sigma_{3}$}

\includegraphics[height=14.0cm,width=15.5cm]{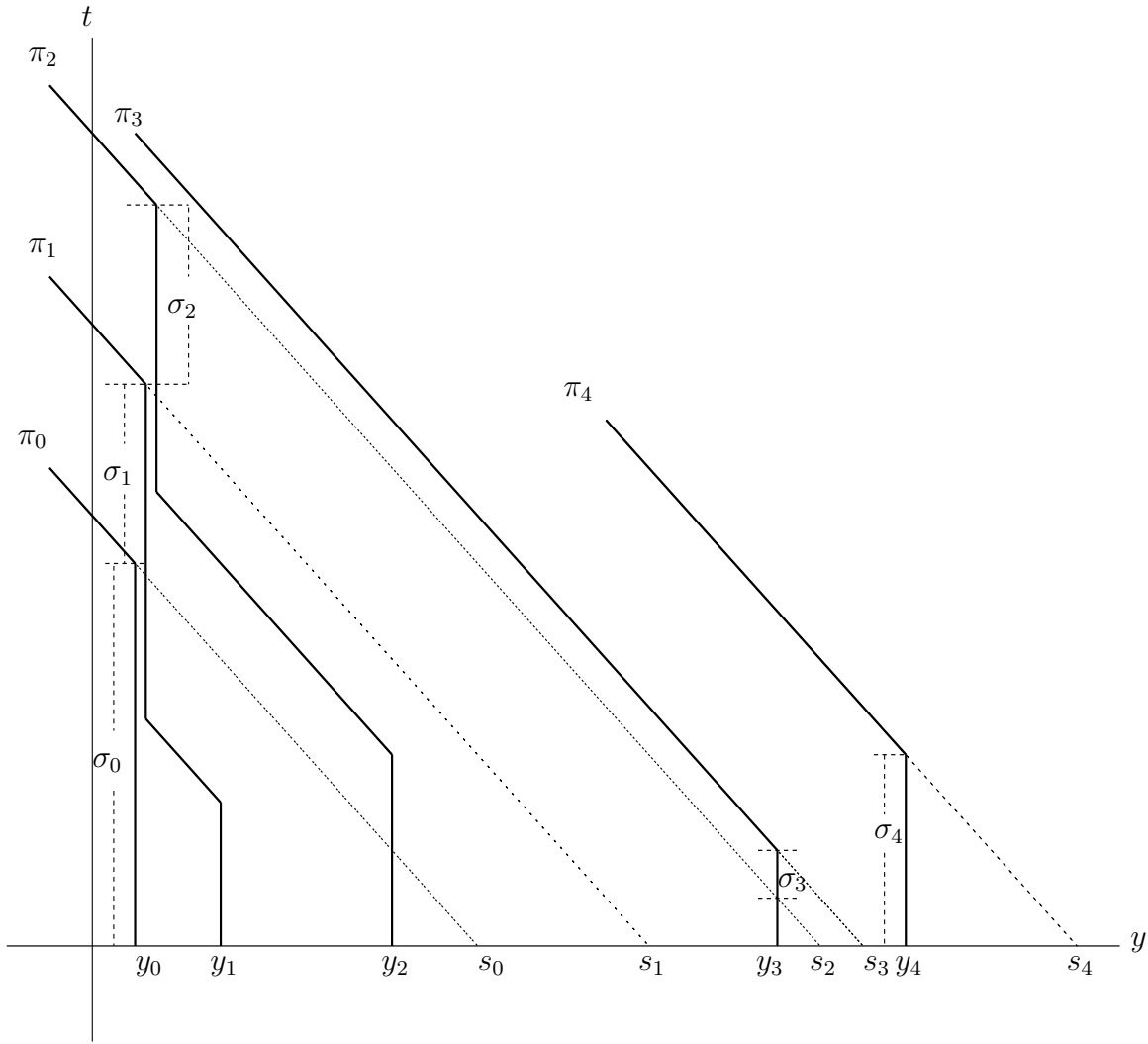}
\caption{The first cycle has 4 cars. Car 1 collides with car 0 and car 2
  collides with car 1 at $y_0$ (the trajectories are drawn slightly separated at
  $y_0$; the actual trajectories partially intersect at $y_0$). Car 3 does not
  collide with previous cars, but since its starting position $y_3$ is to the right of
  $s_2$, it still belongs to the cycle. Car 4 initial position $y_4$ is to the
  right of $s_3$ so that the new cycle starts at $S_1=y_4$.}
\end{figure}

\section{FIFO $M/M/1$ queue}
\label{S4}
The FIFO $M/M/1$ queue is a Markov process in $\N=\{0,1,\dots\}$. At rate
$\lambda>0$ customers arrive into the system, stay in line and one customer at a
time is served at rate 1. This means that the service times are exponentially
distributed with mean 1. The customers respect the time of arrival: \emph{first
  in, first out} in the jargon of queuing theory. We use the following notation

\begin{itemize}
               \item $\ty_i$ arrival time of customer $i$
        \item $\tilde{\sigma}_i$ service time of customer $i$.
        \item $\tilde{s}_i$ exit time of customer $i$.
\end{itemize}

The queue size at time $y$ can be constructed as a function of a marked Poisson
process $((\ty_i,\tilde{\sigma}_i); \, i \in \Z)$. The arrival process $(\ty_i
\,; i\in \Z)$ is a Poisson process of parameter~$\lambda$.  The service times
$(\tsigma_i; \, i\in\Z)$ are \emph{iid} random variables with exponential law of
mean $1$. The sequences are independent.

The \emph{workload} $W_y$, $y \in \R$ is the amount of service due by the server
at time $y$. This is the time a customer arriving at time $y$ has to wait to
start to be served. This process is continuous to the right with limits to the
left and piecewise linear. $W_y$ jumps at times $\ty_i$, the arrival time of
customer $i$ by an amount $\tsigma_i$, the service time of this customer and it
decreases continuously with derivative $-1$ until hitting 0, where stays until
next arrival. The workload process must satisfy the following evolution
equations:
\begin{eqnarray}
  \label{p4}
  \frac {dW_y}{dy} = - \one\{W_y>0\}\qquad \hbox{ for } y\neq \ty_i, \quad i\in\Z\\
  W_{\ty_i} = W_{\ty_i-}+\tsigma_i 
\end{eqnarray}

\paragraph{Construction of the workload process with an arrival to an empty
  system}
Let $\bty = (\ty_0,\ty_1,\dots)$ be a Poisson process of rate $\lambda$ as
$\by$, with $y_0=0$ and let $\btsigma=(\tsigma_0,\tsigma_1,\dots)$ be a sequence
of iid exponential random variables with mean $1$ as $\bsigma$. Define
\begin{eqnarray}
  \label{p13}
  W_0 := \tsigma_0
\end{eqnarray}
and then, recursively, for $i\ge 1$:
\begin{eqnarray}
  \label{p14}
  W_{\ty_i} := [W_{\ty_{i-1}} - (\ty_i-\ty_{i-1})]^+ + \tsigma_i \\
  W_y := [W_{\ty_{i-1}} - (y-\ty_{i-1})]^+ ,\qquad y\in(\ty_{i-1},\ty_i)\label{p14a}
\end{eqnarray}
To stress the dependence of $(\bty,\btsigma)$ we denote $W(\bty,\btsigma)=
(W_y(\bty,\btsigma);\,y\ge 0)$ the process defined by \eqref{p13}, \eqref{p14}
and \eqref{p14a}.

The exit time of customer $i$ is defined by
\begin{equation}
  \label{p20a}
  \tilde{s}_i:=\ty_i+ W_{\ty_i} 
\end{equation}
for $i\geq 0$. That is, the $i$th arrival time plus the workload of the server at
arrival (included the $i$th service time). Define
\begin{equation}
  \label{p20}
  \bts(\bty,\btsigma) := (\ts_0,\ts_1,\dots)
\end{equation}

\paragraph{\bf Workload cycle} 

Let $\tX$ be the first time after time zero the workload jumps from 0 to a
positive value:
\[
\tX := \inf\{y>0:\, W_y- = 0,\, W_y>0\}
\]
Define
\[
\tC := (W_y;\, y\in[0,\tX))
\]
and let $\tN$ be the number of arrivals during the cycle:
\[
\tN := \max\{i:\, \ty_i\in[0,\tX)\} +1
\]
(we add 1 to take account of the arrival at time 0).

Clearly $\tC$, $\tX$ and $\tN$ are function of $(\bty,\btsigma)$:
\[
\tC = \tC(\bty,\btsigma),\quad \tX=\tX(\bty,\btsigma),\quad\tN =\tN
(\bty,\btsigma)
\]

\begin{vthm}
\label{v7}
If $\lambda<1$ then $\tX$ has finite expectation and there exists a stationary
version of the workload process denoted $W=(W_y;\,y\in\R)$.
\end{vthm}
For a proof see Loynes \cite{Loynes}; we prove this later using cycles,
following \cite{Thorisson}. The proof of the following theorem can be found in  
 Prabhu \cite{Prabhu}, theorem $8$ (page 98) or Baccelli and
Br\'emaud \cite{Baccelli}.

\begin{teo}[Burke's Theorem]
\label{burke}
   The exit times $(s_i;\,i\in\Z)$ of the stationary
  process $W$ is a Poisson process with parameter $\lambda$.
\end{teo}

\begin{figure}[!htp]

\psfrag{i1}{$W_y$}
\psfrag{i2}{$y$}
\psfrag{i3}{$y_{0}$}
\psfrag{i4}{$y_{1}$}
\psfrag{i5}{$y_{2}$}
\psfrag{i6}{$\tilde{s}_{0}$}
\psfrag{i7}{$\tilde{s}_{1}$}
\psfrag{i8}{$\tilde{s}_{2}$}
\psfrag{i9}{$y_{4}$}
\psfrag{i10}{$\tilde{s}_{4}$}
\psfrag{i15}{$y_{3}$}
\psfrag{i16}{$\tilde{s}_3$}

\psfrag{i11}{$\tilde{\sigma_{0}}$}
\psfrag{i12}{$\tilde{\sigma_{1}}$}
\psfrag{i13}{$\tilde{\sigma_{2}}$}
\psfrag{i14}{$\tilde{\sigma_{4}}$}
\psfrag{i17}{$\tilde{\sigma_{3}}$}

\includegraphics[height=7.75cm,width=15.5cm]{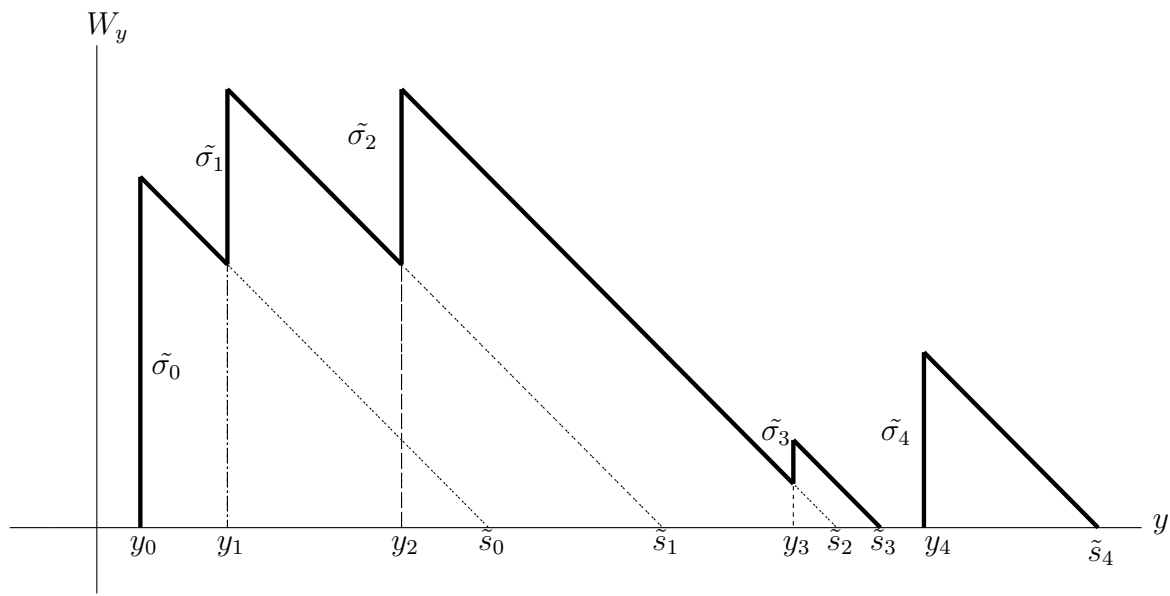}
\caption{Workload evolution related with the traffic example of Figure 1. The
  cycle has 4 customers. The exit time $s_3$ of customer 3 occurs before the
  time arrival $y_4$ of customer~4 so that a new cycle starts at $S_1=y_4$. The
  service times $\tsigma_i$ coincide with the final delays $\sigma_i$ and the
  customer exit times $\ts_i$ coincide with the final relative car positions
  $s_i$ of Figure 1.}
\end{figure}

\section{Queue associated to traffic model} 
\label{S5}

The car trajectories generated by semi-infinite car positions $\by$ and
car-delays $\bxi$ defined in \eqref{p34} generate final car-delays $\bsigma =
\bsigma(\by,\bxi)$ and last passages through the origin $B_{i,0}(\by,\bxi)$. 
The queue generated by arrivals $\by$ and waiting times $\bsigma$ produce a
workload process $W_y(\by,\bsigma)$.
The relation between these processes is given by
\begin{equation}
  \label{p60}
  W_y = (B_{i,0}-y)^+ \qquad \hbox{for }y\in [y_i,y_{i+1})
\end{equation}
This follows from the definitions \eqref{p13}-\eqref{p14a} and \eqref{t2}.  
In the corresponding cycles this relation reads
\begin{equation}
  \label{p61}
  W_y = \left\{\begin{array}{ll} 
      B_{i,0}-y& \mbox{if}\,\ y_i\le y <y_{i+1}\le y_{N-1}, \\
      0 & \mbox{if} \,\ y_{N-1}\le y\le y_N
    \end{array}\right.
\end{equation}
As a consequence of \eqref{p60}, the queue exit times $\bts=\bts(\by,\bsigma)$
coincide with the final relative car-positions $\bs$. The length and number of
elements in the respective cycles agree:
\begin{vthm}
\label{t9}
Let $(\by,\bxi)$ be semi infinite car initial positions and delay times as in
\eqref{p34}. 
Then
\begin{eqnarray}
  \label{p41}
  \bts(\by,\bsigma(\by,\bxi))=\bs(\by,\bxi)\\
  \tX(\by,\bsigma(\by,\bxi))= X(\by,\bxi),\quad \tN(\by,\bsigma(\by,\bxi))=
  N(\by,\bxi)\label{p44}
\end{eqnarray}
\end{vthm}

\paragraph{Proof} \eqref{p44} is a consequence of \eqref{p41} and the
definitions.  It suffices to prove \eqref{p41} for each cycle.
$\sigma_0=\xi_{0,0}=D_0$, so that by \eqref{s_i}, $s_0=0+\sigma_0=\ts_0$ by
\eqref{p20a} because $W_0=\sigma_0$. For $1<i<N$,
\[
\ts_i = y_i + W_{y_i} = B_{i,0} = s_i
\]
by \eqref{p20a}, \eqref{p61} and \eqref{p51}.

\section{Stationary traffic process}
\label{S6}
Let $((\by_n,\bxi_n);\,n\in\Z)$ be a iid sequence of Poisson processes and delay
times with the same law as $(\by,\bxi)$. Let $C_n = C(\by_n,\bxi_n)$ be the
cycles generated by these variables.  Then $(C_n;\,n\in\Z)$ is a sequence of iid
cycles with the same distribution as $C$. Let $X_n$ be the length of cycle
$C_n$, $N_n$ the number of cars involved with this cycle. By \eqref{p44} and
Lemma~{\ref{v7}} $X_n$ has finite expectation. Let
\begin{equation}
  \label{p27}
  S^o_n:=\left\{ \begin{array}{ll}
      0&\hbox{if }n= 0\\
      \sum_{k=1}^n X_k &\hbox{if }n> 0\\
      -\sum_{k=n}^{0} X_k &\hbox{if }n\le0
\end{array}\right.
\qquad
M_n:=\left\{ \begin{array}{ll}
    0&\hbox{if }n= 0\\
    \sum_{k=1}^n N_k &\hbox{if }n> 0\\
    -\sum_{k=n}^{0} N_k &\hbox{if }n\le0
\end{array}\right.
\end{equation}
Define the traffic process $Z^o=(Z^o_y;\,y\in\R)$ by
\begin{equation}
  \label{p23}
  Z^o_y  := Z_{y-S_n}(C_n) \qquad\qquad \hbox{ for } y\in [S^o_n,S^o_{n+1}),\quad n\in\Z
\end{equation}
as the process obtained by juxtaposing the cycles one after the other and
putting the beginning of cycle 0 at the origin; recall the definition of
$Z_y(C)$ in \eqref{p24}.  In the process $Z^o$ the cars of a cycle are
\emph{not} blocked by cars of the previous cycle; in particular, all cars
starting at positions $S_n$, $n\in\Z$, will be free after waiting an exponential
time of mean~1.

Let $U$ be a uniform random variable in $[0,1]$ independent of $Z^o$. Let $\P^o$
be the law of $(((\by_n,\bxi_n);\,n\in\Z),U)$ and $\E^o$ the corresponding
expectation. As $Z^o$ is a function of $((\by_n,\bxi_n);\,n\in\Z)$, we can also
think that $\P^o$ is the law of $(Z^o,U)$. Let $\theta_r$ be the translation
operator defined by $(\theta_r Z)_y = Z_{y-r}$. Define $Z$ as a function of $Z^o$
and $U$ as the process $Z^o$ ``as seen from $-(1-U)X_0$'':
\begin{eqnarray}
  \label{p26}
  Z &:=\theta_{-(1-U)X^o_0} Z^o\qquad y\in\R 
\end{eqnarray}
and define the law $\P$ by size biasing cycle zero:
\begin{equation}
  \label{p28}
  d\P := \frac {X_0}{\E^o X_0} d\P^o
\end{equation}
The following result is in
Theorem 4.1 of Chapter 8 of Thorisson \cite{Thorisson}; see also Figure 2.1 in
that chapter for a better understanding of the relation between $Z$ and $Z^o$. 

\begin{vthm}
  \label{p30a}
The law of $Z$ under $\P$ is stationary. 
\end{vthm}

For test functions $f$ (from the space where $Z$ is defined to $\R$) the law of
$Z$ under $\P$ satisfies
\[
\E f(Z) = \frac {\E^o [X_0f(\theta_{-X_0(1-U)}Z^o)] }{\E^o X_0}  
\]
where $X_0$ is the lenght of cycle 0 of $Z^o$.  To obtain a sample of $Z$ under
$\P$, first sample a process with a cycle starting at the origin with the size
biased law $\P$, then translate the origin to a point uniformly distributed in
cycle 0.
 
Since the traffic process $Z$ is constructed as juxtaposition of cycles, the
initial car positions, the final delays and the final relative car positions can
be recovered as follows using the notation of \eqref{p38}:
\begin{eqnarray}
  \label{p37}
  S_0 = \sup\{y\le 0\,:\,Z_{y-}=0,\, Z_{y}>0\}, \\
  S_n = \inf\{y>S_{n-1}\,:\, Z_{y-}=0,\, Z_{y}>0\}, n>0,\\
  S_n = \sup\{y<S_{n+1}\,:\, Z_{y-}=0,\, Z_{y}>0\}, n<0\,;\\
  X_n = S_{n+1}-S_n,\quad C_n = (Z_{y-S_n};\, y\in[S_n,S_n+X_n)),\quad N_n=N(C_n)\,;\\
  y_k = S_n + y_{k-M_n}(C_n) ,\quad \sigma_k =  \sigma_{k-M_n}(C_n),
  \quad s_k = S_n + s_{k-M_n}(C_n),\\
  \qquad\qquad \qquad \qquad\qquad \qquad \qquad  \qquad \hbox{ for } M_n\le k < M_{n+1}.\label{p40}
\end{eqnarray}
Denote $\by(Z)$, $\bsigma(Z)$ and $\bs(Z)$ the initial car-positions, the final
delays and final relative car-positions of the process $Z$.

\begin{vthm}
  \label{p30}
  The law of $\by(Z)$ and $\bsigma(Z)$ under $\P$ are stationary and $\by(Z)$
  and $\bsigma(Z)$ are independent. In particular $\by(Z)$ is a Poisson process
  of parameter $\lambda$ and $\bsigma(Z)$ is a sequence of iid exponential
  random variables of mean 1. Furthermore $\bs(Z)$ is a Poisson process of
  parameter $\lambda$.
\end{vthm}

Stationarity of $\by(Z)$ and $\bsigma(Z)$ is immediate consequence of
stationarity of $Z$. The other properties follow from the stationary
construction of the queue and relation \eqref{p41} as shown in the next section.

\paragraph{Trajectory extension} The trajectory of the car starting
at $y_i\in[S_n,S_{n+1})$ is defined only for $t\in[0,B_{i,0}(C_n)]$; see
\eqref{p24}. We extend the trajectory by just continuing at speed $1$ from this
point on:
\begin{equation}
  \label{pi}
  \pi_i(t) = 
\left\{
\begin{array}{ll}
S_n+ \pi^n_{i-M_n}(t)& \mbox{if}\,\ y_i\in[S_n,S_{n+1}) , \; t\in[0,B_{i,0}(C_n)] \\
S_n-t+ B_{i-M_n,0}(C_n) & \mbox{if} \,\ y_i\in[S_n,S_{n+1}) , \; t>B_{i,0}(C_n)
\end{array}
\right.
\end{equation}
where $\pi^n_i$ is the trajectory of the $i$th particle of cycle $C_n$.  Using
definitions \eqref{pi} and \eqref{p24} the corresponding process $Z$ has a
countable number of coordinates at each position $y$; the $k$th coordinate
indicates the interval of time spent at $y$ by the $k$th car initially to the
right of $y$. We abuse notation and continue calling $Z$ this process. Notice
that $\by(Z)$, $\bsigma(Z)$ and $\bs(Z)$ remain unchanged and Lemma \ref{p30}
holds for this extension.

\section{Stationary workload process}
\label{S7}
Assume $\lambda<1$ and let $(\by_n,\bxi_n)$ be the sequence introduced in the
beginning of previous section. Let $\bsigma_n=\bsigma(\by_n,\bxi_n)$ as defined
in \eqref{t2}.  By \eqref{p44} the workload cycles $\tC_n = \tC(\by_n,\bsigma_n)$
have the same length as the traffic cycles $C_n$: $\tX_n=X_n$ and the number of
cars in cycle $C_n$ is the same as the number of customers in cycle $\tC_n$:
$\tN_n=N_n$.  Let $S_n$ as in \eqref{p27} and define $W^o=(W^o_y;\,y\in\R)$ by
\begin{equation}
  \label{p31}
  W^o_y  = W_{y-S_n}(\tC_n) \qquad\qquad \hbox{ for } y\in [S^o_n,S^o_{n+1}),\quad n\in\Z
\end{equation}
and the process $W=(W_y;\,y\in\R)$ by
\begin{equation}
  \label{p32}
    W_y = W^o_{y+(1-U)X^o_0}\qquad y\in\R
\end{equation}
where $U$ is the same variable used in \eqref{p26}. As before, for $\P$ given by
\eqref{p28},
\begin{vthm}
  \label{p33a}
Under $\P$ the law of $W$ is stationary. 
\end{vthm}

$W$ is a stationary $M/M/1$ queue with arrivals $\by(W)$ and departures
$\bts(W)$. Hence the arrival process $\by(W)$ is a stationary Poisson process of
rate $\lambda$ in $\R$. By Burke's Theorem {\ref{burke}}, the same is true
for the departure process $\bts(W)$.

The stationary workload process $W$ and the stationary traffic process $Z$ are
constructed in the same space (as function of $((\by_n,\bxi_n);\,n\in\Z),U)$ so
that they have exactly the same cycles and the initial and relative final car
positions of $Z$ coincide respectively with the arrival and departure process of $W$:
\begin{equation}
  \label{p47}
  \by(Z) = \by(W), \quad \bs(Z)=\bts(W)
\end{equation}
This finishes the proof of Lemma {\ref{p30}} and Proposition {\ref{t6a}}.

\section{Final remarks}

When the initial density of cars is smaller than the inverse of the delay time,
$\lambda<1$ the process is called \emph{subcritical}.  We have described how a
stationary configuration of initial car positions organize the departure from
speed zero to speed $1$ under the rule slow-to-start in the subcritical
case. The method relates the space-stationary one-dimensional slow-to-start
traffic model with the workload process of a $M/M/1$ time-stationary queuing
system. If the initial position of cars is Poisson then the final relative
position of free cars is also Poisson. The same method shows that if the initial
position of cars $\by$ is a stationary ergodic process with density $\lambda<1$
(density is the mean number of cars per unit length) then the final relative car
position process coincides with the departure process of the queue with arrival
process~$\by$.

The supercritical case merits to be investigated. When $\lambda>1$ cars form
``traffic jams'' on a subset of the initial positions $\by$, depending on
time. As time grows, the density of this set goes to zero and the number of cars
per traffic jam increases. We agree with an anonymous referee who conjectures
that as $t$ goes to infinity the set of traffic jams suitably rescaled converges
to a Poisson process of rate 1. The critical case $\lambda=1$ should also show
traffic jams.

Another model to research might include spontaneous stops at some rate $\nu$
keeping the slow-to-start rule. These kind of models will be closer to those
studied by Nagel and Schreckenberg and Gray and Griffeath.

We are investigating the same phenomena for cellular automata in $\Z$. The
approach can be applied but there are complications coming from the hard core
interaction. The results are similar.

\section*{Acknowledgements}
We thank Andreas Schadschneider and David Griffeath for discussions about
slow-to-start models and Luis Renato Fontes for his inspiring comments. We thank
the referees for their reading and comments. 

This paper is partially supported by Funda\c c\~ao de Amparo \`a Pesquisa do
Estado de S\`ao Paulo FAPESP, Funda\c c\~ao de Amparo \`a Pesquisa do Estado de
Minas Gerais FAPEMIG, Programa N\'ucleos de Excel\^encia PRONEX, Conselho
Nacional de Desenvolvimento Cient\'{\i}fico e Tecnol\'ogico CNPq and Instituto
do Mil\^enio Avan\c co Global e Integrado da Matem\'atica Brasileira, IM-AGIMB.
E. P. was partially support by CRDF, Grant RUMI-2693-MO-05.  Part of this work
was done while the P.A.F. was visiting Isaac Newton Institute for Mathematical
Sciences during the program Principles of the Dynamics of Non-Equilibrium
Systems in 2006. Hospitality and support is greatly acknowledged.

\end{document}